\documentclass[english,showpacs,floatfix,11pt]{revtex4}
\usepackage[dvips]{graphicx}
\usepackage{longtable}
\usepackage{amsmath,amssymb}
\usepackage{dcolumn}
\usepackage{latexsym}
\usepackage{babel}
\usepackage[latin1]{inputenc}
\usepackage{color}
\usepackage[colorlinks]{hyperref}
\def\al{\alpha}

\def\nb{\nabla}
\def\pa{\partial}
\def\vf{\varphi}
\def\vr{\varrho}

\def\eps{\epsilon}
\def\Up{\Upsilon}
\def\Om{\Omega}
\def\om{\omega}
\def\Ga{\Gamma}
\def\ga{\gamma}
\def\be{\beta}

\def\dl{\delta}

\def\th{\theta}
\def\sg{\sigma}
\def\wt{\widetilde}
\def\H{\textbf{H}}
\def\l{\left}
\def\r{\right}
\def\nn{\nonumber}
\def\diag{\mbox {diag}}

\begin{document}
\title{\Large\bf The Simplification of Spinor Connection and Classical Approximation}
\author{Ying-Qiu Gu}
\email{yqgu@fudan.edu.cn} \affiliation{School of Mathematical
Science, Fudan University, Shanghai 200433, China} 
\date{25th November 2017}%\today}%28th September 2017}%

\begin{abstract}
The standard spinor connection in curved space-time is represented
in a compact form. In this form the calculation is complicated, and
its physical effects are concealed. In this paper, we split spinor
connection into two vectors $\Upsilon_\mu$ and $\Omega_\mu$, where
$\Upsilon_\mu$ is only related to geometrical calculations, but
$\Omega_\mu$ leads to dynamical effects, which couples with the spin
of a spinor. The representation depends only on metric but is
independent of Dirac matrices, so it is valid for both  Weyl spinors
and Dirac spinor. In the new form, we can clearly define classical
concepts for a spinor and then derive its complete classical
dynamics. By detailed calculation we find the classical
approximation is just Newtonian second law. The dynamical connection
$\Omega_\mu$ couples with the spin of a particle with a tiny energy
in weak field, which provides location and navigation functions for
a spinor. This term may be also important to form magnetic field of
a celestial star. From the results, we find the spinor has
marvelous structure and wonderful property, and the interaction
between spinor and gravity is subtle. This study may be also helpful
to clarify the relations between relativity, quantum mechanics and
classical mechanics.

\end{abstract}
\maketitle

\section{Introduction}

The classical theory of motion for a spinor in a gravitational field
is firstly studied by Mathisson\cite{sp1}, and then developed by
Papapetrou\cite{sp2} and Dixon\cite{sp3}. A detailed derivation can
be found in \cite{FFF}. Where by the commutator of the usual
covariant derivative of the spinor $[\nb_{\al}, \nb_{\be}]$, we get
an extra approximate acceleration of the spinor as follows
\begin{eqnarray}
a_\al(x^\mu) = -\frac{\hbar}{4 m} R_{\al\be\ga\dl}(x^\mu)
u^\be(x^\mu) S^{\ga\dl}(x^\mu), \label{acc}\end{eqnarray} where
$R_{\al\be\ga\dl}$ is the Riemann curvature, $u^\al$ 4-vector speed
and $S^{\ga\dl}$ the half commutator of the Dirac matrices.

(\ref{acc}) leads to the violation of Einstein's equivalence
principle. This problem was discussed by many
authors\cite{FFF,spin1,spin2,spin3,spin4,spin5,spin6,spin7}. In
\cite{spin1}, the exact Cini-Touschek transformation and the
ultra-relativistic limit of the fermion theory were derived, but the
Foldy-Wouthuysen transformation is not uniquely defined. The
following calculations also show that, the usual covariant
derivative $\nb_\mu$ includes cross terms, which is not parallel to
the speed $u^\mu$ of the spinor.

{For a classical spin such as gyroscope, the frame dragging effect
was predicted by Lense and Thirring \cite{LT1,LT2}, and the
non-relativistic formula for the effect was derived by L. Schiff
\cite{Schiff1,Schiff2,Schiff3}. It has also been shown that the
gravitomagnetic interaction plays a part in both shaping the lunar
orbit\cite{EEP9}, and in contributing to the periastron precession
of binary and especially double pulsars\cite{gyro2}. For
applications to the analysis of gravitational phenomena, a general
metric tensor field expansion for the gravitational potentials in a
broad class of theories was developed\cite{gyro4,gm1,gm2,gm4}. This
parameterized post-Newtonian framework yields a gravitomagnetic
contribution to the equation of motion\cite{gyro5}. The spin
precession was studied in \cite{prec}.

In this paper, by projecting the spinor connection onto the tetrad
or Pauli matrices, and splitting it into geometrical and dynamical
parts, we get two 4-d vectors $(\Up_\mu, \Om_\mu)$ from the
connection. These vectors of connection are only determined by
metric but independent of Dirac matrices, and the classical
approximation is parallel to 4-vector speed of particle. In this
representation of connection, we can clearly define classical
concepts such as coordinate, speed, momentum for a spinor, and then
derive the classical mechanics in detail. $\Up_\mu$ only corresponds
to the geometrical calculations, but $\Om_\mu$ leads to tiny
dynamical effects. $\Om_\mu$ couples with the spin $s^\mu$ of a
spinor, which provides location and navigation functions for a
spinor with little energy. So this form of connection is helpful to
understand the subtle interaction between spinor and gravity.

\section{Simplification of the spinor connection}
\setcounter{equation}{0} At first we introduce some notations and
conventions. We take $\hbar=c=1$ as units, the Minkowski metric is
given by $\eta_{ab}=\diag(1,-1,-1,-1)$, the Pauli and Dirac matrices
in Minkowski space-time is as follows
\begin{eqnarray}
 \sg^{a}\equiv \left \{\left(\begin{array}{cc}
 1 & 0 \\  0 & 1\end{array}\right),\left(\begin{array}{cc}
 0 & 1 \\  1 & 0\end{array}\right),\left(\begin{array}{cc}
 0 & -i \\  i & 0\end{array}\right),\left(\begin{array}{cc}
 1 & 0 \\  0 & -1\end{array}\right)
 \right\},
\label{1.1}\end{eqnarray}
\begin{eqnarray}
\wt{\sg}^a \equiv(\sg^0,-\vec\sg),\qquad
\vec\sg=(\sg^1,\sg^2,\sg^3). \label{1.2} \end{eqnarray}
\begin{eqnarray}\begin{array}{l}
\ga^a\equiv \left(\begin{array}{cc}0 & \wt \sg^a\\ \sg^a &
0\end{array}\right),\qquad \ga_4=\left(\begin{array}{cc} I & 0\\ 0&
-I
\end{array}\right).\end{array} \label{1.3}
\end{eqnarray}
The element of space-time is described by
\begin{eqnarray}
d\mathbf{x}=\wt\ga_\mu dx^\mu=\ga_a \dl X^a, \quad \wt\ga^\mu =
h^\mu_{~a} \ga^a,\quad \wt\ga_\mu = l_\mu^{~a} \ga_a,
\label{dx}\end{eqnarray} in which  $\ga_a$ and $\wt\ga_\mu$ act as
tetrad frames satisfying  the following $C\ell({1,3})$ Clifford
algebra,
\begin{eqnarray}
\ga_a \ga_b+\ga_b\ga_a =2\eta_{ab},\qquad
\wt\ga_\mu\wt\ga_\nu+\wt\ga_\nu\wt\ga_\mu=2g_{\mu\nu}. \label{1.2a}
\end{eqnarray}
In this paper, we use the indices $(a,b\in\{0,1,2,3\})$ for the
Minkowski space-time, Greek characters $(\mu,\nu\in\{0,1,2,3\})$ for
the curved space-time, and $(j,k,l\in\{1,2,3\})$ for the
simultaneous hypersurface or space.

In the flat space-time, the Dirac equation for free bispinor $\phi$
is equivalent to
\begin{eqnarray}
\ga^a i\pa_a\phi=m\phi. \label{1.4}
 \end{eqnarray}
In chiral representation we get dynamics of Weyl spinors,
\begin{eqnarray}\left\{\begin{array}{l}
\sg^a i\pa_a\psi=m\wt\psi,\\
\wt\sg^a i\pa_a\wt\psi=m\psi,
\end{array}\right.\qquad \phi=\left(\begin{array}{c}\psi \\\wt\psi \end{array}\right), \label{1.6} \end{eqnarray}
which is more convenient for calculation than (\ref{1.4}) in some
cases.

In curved space-time we have  Pauli and Dirac matrices as follows
\begin{eqnarray}\left\{\begin{array}{l}
\vr^\mu=h^\mu_{~a}\sg^a,\quad\vr_\mu=l_\mu^{~a}\sg_a, \\
\wt\vr^\mu=h^\mu_{~a}\wt\sg^a, \quad \wt\vr_\mu=l_\mu^{~a}\wt\sg_a,
\end{array}\right.\qquad \wt\ga^\mu = \left(\begin{array}{cc} 0 & \wt \vr^\mu \\ \vr^\mu & 0
\end{array}\right).\label{1.9}
\end{eqnarray}
The spinor equation (\ref{1.6}) becomes
\begin{eqnarray}\left\{\begin{array}{l}
\vr^\mu i\nb_\mu\psi=m\wt\psi,\\
\wt\vr^\mu i\wt\nb_\mu\wt\psi=m\psi,
\end{array}\right. \label{1.10} \end{eqnarray}
where $\nb_\mu=\pa_\mu+\Ga_\mu$,$~\wt\nb_\mu=\pa_\mu+\wt\Ga_\mu$ are
the covariant derivatives of $\psi$ and $\wt\psi$, $\Ga_\mu$ and
$\wt\Ga_\mu$ are spinor affine connections\cite{1,2,3,4,5},
\begin{eqnarray}
\Ga_\mu=\frac 1 4 \wt\vr_\nu\vr^\nu_{;\mu},\qquad \wt\Ga_\mu=\frac 1
4 \vr_\nu\wt\vr^\nu_{;\mu}, \label{1.11}\end{eqnarray} in which
$\vr^\mu_{;\nu}=\pa_\nu\vr^\mu+\Ga^\mu_{\al\nu}\vr^\al$. For Dirac
bispinor $\phi$, by (\ref{1.11}) it is easy to check
\begin{eqnarray}
\nb_\mu\phi=(\pa_\mu+\hat\Ga_\mu)\phi,\qquad \hat\Ga_\mu=\frac 1 4
\wt\ga_\nu\wt\ga^\nu_{;\mu}. \label{1.12}
\end{eqnarray}

The Lagrangian corresponding to (\ref{1.10}) is given by
\begin{eqnarray}
{\cal L}_m &=& \Re \l(\psi^+\vr^\mu i\nb_\mu\psi+\wt\psi^+\wt\vr^\mu
i\wt\nb_\mu\wt\psi\r)-m(\wt\psi^+\psi+\psi^+\wt\psi),\nn\\
&=&\Re  \l(\psi^+\vr^\mu i\pa_\mu\psi+\wt\psi^+\wt\vr^\mu
i\pa_\mu\wt\psi \r)+\psi^+\Om \psi+\wt\psi^+\wt\Om
\wt\psi-m(\wt\psi^+\psi+\psi^+\wt\psi), \label{2.2}\end{eqnarray} in
which $\Om$ and $\wt\Om$ are two Hermitian matrix defined by
\begin{eqnarray}\left\{\begin{array}{l}
\Om\equiv  \frac i 8[\vr^\mu
\wt\vr^\al\pa_\mu\vr_\al-(\pa_\mu\vr_\al)\wt\vr^\al\vr^\mu],\\
\wt\Om\equiv  \frac i 8[\wt\vr^\mu
\vr^\al\pa_\mu\wt\vr_\al-(\pa_\mu\wt\vr_\al)\vr^\al\wt\vr^\mu].
\end{array}\right.
\end{eqnarray}
For any diagonal metric, it easy to check $\Om=\wt\Om=0$. By
variation of (\ref{2.2}) with respect to $\psi^+$ and $\wt\psi^+$,
we get dynamics equivalent to (\ref{1.10}) as follows
\begin{eqnarray} \left\{\begin{array}{l}
\vr^\mu i\pa_\mu\psi+(\frac i 2 \vr^\mu_{;\mu}+\Om)\psi=m\wt\psi,\\
\wt\vr^\mu i\pa_\mu\wt\psi+(\frac i 2
\wt\vr^\mu_{;\mu}+\wt\Om)\wt\psi=m\psi.
\end{array}\right.
\label{2.3} \end{eqnarray}

Projecting $\pa_\mu\vr^\mu$ onto the basis $\vr^\mu$, i.e. we define
$k_\mu$ as follows
\begin{eqnarray}
\pa_\mu\vr^\mu=\pa_\mu h^\mu_{~a}\sg^a \equiv k_\mu \vr^\mu=k_\mu
h^\mu_{~a}\sg^a,\end{eqnarray} then we have $\pa_\mu
h^\mu_{~a}=k_\mu h^\mu_{~a}$ or $k_\mu=l_\mu^{~a}\pa_\nu
h^\nu_{~a}$, and
\begin{eqnarray}
\vr^\mu_{;\mu}=\pa_\mu\vr^\mu+\Ga^\mu_{\mu\nu}\vr^\nu=(l_\mu^a\pa_\nu
h^\nu_a+\pa_\mu\ln\sqrt{g})\vr^\mu. \label{2.4}\end{eqnarray} So we
can define the geometrical part of connection by
\begin{eqnarray}
\Up=\Up_\mu\vr^\mu\equiv \frac1 2 \vr^\mu_{;\mu},\quad
\Upsilon_\mu\equiv \frac 1 2 (l_\mu^a\pa_\nu
h^\nu_a+\pa_\mu\ln\sqrt{g})= \frac 1 2 h^\nu_a(\pa_\mu
l_\nu^a-\pa_\nu l_\mu^a). \label{2.5}
\end{eqnarray}

For any 3-d vectors $\vec A$ and $\vec B$, we have
\begin{eqnarray}
(\vec A\cdot \vec \sg)(\vec B\cdot \vec\sg)=\vec A\cdot\vec B+i(\vec
A\times\vec B)\cdot \vec\sg.\end{eqnarray} Denoting
\begin{eqnarray}
\vr^\al=h^\al_{~0}+\vec h^\al\cdot \vec
\sg,~~\wt\vr^\al=h^\al_{~0}-\vec h^\al\cdot \vec
\sg,~~\pa_\mu\vr_\al=\pa_\mu l_\al^{~0}-\pa_\mu\vec l_\al\cdot \vec
\sg,
\end{eqnarray}
where $\vec h^\al=(h^\al_{~1},h^\al_{~2},h^\al_{~3})$ and $\vec
l_\al=(l_\al^{~1},l_\al^{~2},l_\al^{~3})$, by straightforward
calculation we get
\begin{eqnarray}\Om=-\frac 1 4\left(
(\vec h^\al\times\vec h^\be)\cdot\pa_\al \vec l_\be-\pa_\al
l_\be^{~0}(\vec h^\al\times\vec h^\be)\cdot\vec\sg+[(h^\al_{~0}\vec
h^\be-h^\be_{~0}\vec h^\al)\times\pa_\al\vec
l_\be]\cdot\vec\sg\right).
\end{eqnarray}
Let $\Om=\om_a\sg^a=\Om_\mu\vr^\mu$, then we have
\begin{eqnarray}\left\{\begin{array}{l}
\om_0 = -\frac 1 4 (\vec h^\al\times\vec h^\be)\cdot\pa_\al \vec
l_\be,\\
\vec\om =-\frac 1 4 \l(\pa_\al l_\be^{~0}(\vec h^\al\times\vec
h^\be)-(h^\al_{~0}\vec h^\be-h^\be_{~0}\vec
h^\al)\times\pa_\al\vec l_\be \right),\\
\Om_\mu = -\frac 1 4 \left((\vec h^\al\times\vec
h^\be)\cdot(l_\mu^{~0}\pa_\al \vec l_\be-\vec l_\mu\pa_\al
l_\be^{~0})+\vec l_\mu\cdot[(h^\al_{~0}\vec h^\be-h^\be_{~0}\vec
h^\al)\times\pa_\al\vec l_\be]\right),
\end{array}\right.
\label{2.11}
\end{eqnarray} where $\vec \om=(\om^1,\om^2,\om^3)$. In \cite{eng} we get $(\Om^\al,\om^a)$ expressed by $\pa_\al
g_{\mu\nu}$ as follows,
\begin{eqnarray}
\om^d = \frac 1 8 \eps^{dabc} h^\al_{~a}S^{\mu\nu}_{bc}\pa_\al
g_{\mu\nu},\qquad \Om^\al = \frac 1 8 \eps^{dabc}
l_d^{~\al}h^\be_{~a}S^{\mu\nu}_{bc}\pa_\be g_{\mu\nu}.
 \label{omOm}
\end{eqnarray}
(\ref{2.11}) or (\ref{omOm}) defines the dynamical part of the
spinor connection.

Similarly we have
\begin{eqnarray}
\wt\Up=\Up_\mu\wt\vr^\mu=\frac 1 2\wt\vr^{\mu}_{;\mu},\quad
\wt\Om\equiv \wt\Om_\mu\wt\vr^\mu=-\om_a\wt\sg^a,\quad
\wt\Om_\mu=-\Om_\mu. \label{wtom}
\end{eqnarray}
By (\ref{2.5}) and (\ref{wtom}), the dynamical equation (\ref{2.3})
becomes
\begin{eqnarray} \left\{\begin{array}{l}
\vr^\mu [i(\pa_\mu+\Upsilon_\mu)+\Om_\mu]\psi=m\wt\psi,\\
\wt\vr^\mu [i(\pa_\mu+\Upsilon_\mu)-\Om_\mu]\wt\psi=m\psi.
\end{array}\right.
\label{2.6} \end{eqnarray} Correspondingly, the Dirac equation
(\ref{1.4}) in the curved space-time becomes
\begin{eqnarray}
\wt\ga^\mu[ i(\pa_\mu+\Upsilon_\mu)+\Om_\mu \ga_4]\phi=m\phi.
\label{2.7}
\end{eqnarray}

In order to characterize the rotational degrees of freedom, the
decomposition of spinor connection in Clifford algebra was derived
by J. M. Nester  as follow\cite{nst,dmh},
\begin{eqnarray}
\wt\ga^\mu\nb_\mu\phi=\wt\ga^\mu\pa_\mu \phi-\frac 1 2\wt\ga^\mu \wt
q_\mu \phi +\frac 1 {2\cdot 3!}\hat q_{\mu\nu\om}\wt \ga^{\mu\nu\om}
\phi. \label{nst}
\end{eqnarray}
Simplifying the grade-3 Clifford algebra by
$\ga^{abc}=\eps^{abcd}\ga_d\ga^{0123}$ and combining like terms, we
find the two splits (\ref{2.7}) and (\ref{nst}) are equivalent.
However, (\ref{2.6}) and (\ref{2.7}) are more convenient for
discussion and practical calculation as shown below.

The vector connection $\Upsilon_\mu$ and $\Om_\mu$ are only
determined by metric and get rid of the influence of coefficient
matrices. The following discussion shows that, $\Upsilon_\mu$ and
$\Om_\mu$ have different physical meanings. $\pa_\mu+\Upsilon_\mu$
as a whole operator is similar to the covariant derivatives
$\nb_\mu$ for tensors, it only has geometrical effect. But $\Om_\mu$
couples with the spin of a particle, and leads to dynamical effects.

We calculate Dirac equation in diagonal metric. In general case, the
metric is given by
\begin{eqnarray}g_{\mu\nu}={\rm diag}(N_0^2,-N_1^2,-N_2^2,-N_3^2),\qquad \sqrt{g}=N_0N_1 N_2N_3, \label{3.6}\end{eqnarray}
where $N_\mu=N_\mu(x^\al)$. Then we have $\Om_\mu=0$, and
\begin{eqnarray}
\wt\ga^\mu=\left(\frac {\ga^0}{N_0},\frac{\ga^1}{N_1}, \frac
{\ga^2}{N_2},\frac {\ga^3}{N_3}\right),\quad \Upsilon_k= \frac 1 2
\pa_k \ln \left( \frac{\sqrt{g}} {N_k} \right),  \label{3.8}
\end{eqnarray}
where $k=0,1,2,3$.

For Dirac equation in Schwarzschild metric,
\begin{eqnarray}
g_{\mu\nu}=\diag(B(r),-A(r),-r^2,-r^2\sin^2\th),
\label{3.1}\end{eqnarray} we have
\begin{eqnarray}
\wt\ga^\mu=\left(\frac {\ga^0} {\sqrt{B}},\frac {\ga^1} {\sqrt{A}},
\frac {\ga^2} r ,\frac {\ga^3} {r\sin\th} \right),\quad
\Upsilon_\mu=\left(1,\frac 1 r+ \frac {B'}{4B},\frac 1 2
\cot\th,0\right).\label{3.3}
\end{eqnarray}
The Dirac equation for free spinor is given by
\begin{eqnarray} i\left[\frac {\ga^0} {\sqrt{B}}\pa_t+\frac {\ga^1} {\sqrt{A}}
\left(\pa_r+\frac 1 r+ \frac {B'}{4B}\right)+\frac {\ga^2} r
(\pa_\th+\frac 1 2 \cot\th)+\frac {\ga^3}
{r\sin\th}\pa_\vf\right]\phi=m\phi.
 \label{3.4} \end{eqnarray}
Set $A=B=1$, we  get  Dirac equation  in  spherical coordinate
system
\begin{eqnarray}
i \left[{\ga^0}\pa_t+{\ga^1}(\pa_{r}+
\frac{1}{r})+\frac{\ga^2}{r}(\pa_{\theta}+\frac{1}{2
}\cot\theta)+\frac{\ga^3}{r \sin\theta}\pa_{\varphi}\right]\phi=m
\phi. \label{3.5}\end{eqnarray}

\section{The classical approximation of Dirac equation}
\setcounter{equation}{0}

In this section, we derive the classical mechanics of a spinor
moving in gravity, and disclose the physical meaning of connection
$\Up_\mu$ and $\Om_\mu$. We introduce the local Gaussian normal
coordinate system(GCS) with metric $\diag(1,-\bar g_{jk})$, because
only in such coordinate system we can define simultaneity and then
clearly establish the Hamiltonian formalism and calculate the
N\"other charges. In GCS, we have
\begin{eqnarray}
h^0_{~0}=l^{~0}_0=1,\qquad \vec h^0=\vec l_0=0. \end{eqnarray} Then
by (\ref{2.5}) we get
\begin{eqnarray}
\Up_\mu&=& \frac 1 2 \l(\pa_t\ln\sqrt{g}, ~\vec l_k \cdot \pa_j\vec
h^j+\pa_k\ln\sqrt{g}\r). \label{4.2}
\end{eqnarray}
In GCS, to lift and lower the index of a vector means $\Up^0=\Up_0,
\Up^k=-\bar g^{kl}\Up_l$.

More generally, we consider   Dirac equation with electromagnetic
potential $eA^\mu$, then (\ref{2.7}) can be rewritten in Hamiltonian
formalism
\begin{eqnarray} i(\pa_t+\Up_t)\phi=\H\phi, \label{4.3*}
\end{eqnarray}
where the Hamiltonian is defined by
\begin{eqnarray}
\H= -\al^k \hat p_k+eA_0+m \ga_0-\Om_\mu \hat s^\mu,\qquad
\al^\mu\equiv \ga_0
\wt\ga^\mu=\diag(\vr^\mu,\wt\vr^\mu),\label{haml}
\end{eqnarray}
in which $\al^\mu$ is current operator, and $\hat p_\mu$ and $\hat
s^\mu$ are respectively momentum and spin operators defined by
\begin{eqnarray}
\hat p_\mu= i(\pa_\mu+\Up_\mu)-e A_\mu,\qquad \hat s^\mu\equiv
\al^\mu \ga_4=\diag(\vr^\mu,-\wt\vr^\mu).\label{ps}
\end{eqnarray}
It is easy to check $\vec {\hat s}=\diag (\vec \sg, \vec \sg)$ is
the usual spin for any representation of Dirac bispinor.

Similarly to the case in flat space-time\cite{gu2,mass}, we define
  classical concepts such as coordinate $\vec X$ and speed $\vec
v$ of the spinor as follows,
\begin{eqnarray}
\vec X(t)= \int_{S^3} \vec x q^0\sqrt{g} d^3x,\qquad \vec v=\frac d
{dt} \vec X,\label{4.4}
\end{eqnarray}
where $S^3$ stands for the total simultaneous hypersurface, $q^\mu$
is   current
\begin{eqnarray}
q^\mu=\phi^+\al^\mu \phi=\psi^+\rho^\mu \psi +\wt \psi^+\wt \rho^\mu
\wt \psi.\label{4.5}
\end{eqnarray}
By   definition (\ref{4.4}) and   current conservation law
$q^\mu_{;\mu}=0$, it is easy to check
\begin{eqnarray}
\vec v = \int_{S^3} \vec x \pa_t(q^0 \sqrt g)  d^3x=\int_{S^3} \vec
x q^0_{;t} \sqrt g  d^3x=-\int_{S^3} \vec x q^k_{;k} \sqrt g
d^3x=\int_{S^3}\vec q\sqrt{g} d^3x. 
\end{eqnarray}
With   normalizing condition $\int_{S^3}q^0\sqrt{g} d^3x=1$, we have
  point-particle model,
\begin{eqnarray}
q^\mu\to u^\mu \sqrt{1-\bar g_{kl}v^k v^l}\dl^3(\vec x-\vec
X),\qquad u^\mu\equiv \frac {d X^\mu}{d\tau}=(1, \vec v)/
\sqrt{1-\bar g_{kl}v^k v^l}, \label{4.6*}
\end{eqnarray}
where the Dirac-$\dl$ means $\int_{S^3}\dl^3(\vec x-\vec X)\sqrt{g}
d^3x=1$ and $\dl^3(\vec x-\vec X)=0$ if $\vec x\ne\vec X$, $\tau$ is
  proper time $d\tau = \sqrt{1-\bar g_{kl}v^k v^l } dt$.

For any Hermitian  operator $\hat P$, by (\ref{4.3*}) we have
following generalized Ehrenfest theorem,
\begin{eqnarray}
\frac {dP}{dt}&=&\frac d {dt} \int_{S^3}\sqrt g  \phi^+\hat P\phi d^3x \nn\\
&=& \Re\int_{S^3}\sqrt g  \l(\phi^+ (\pa_t\hat P)\phi+
i(i\pa_t\phi)^+ \hat P\phi-i\phi^+ \hat P(i\pa_t\phi)+\phi^+ \hat
P\phi\pa_t\ln\sqrt g \r)d^3x, \nn\\&=& \Re\int_{S^3}\sqrt g
\l(\phi^+ (\pa_t\hat P)\phi+ i(\H\phi)^+ \hat P\phi-i\phi^+ \hat
P\H\phi\r) d^3x,\nn\\
&=&\Re\int_{S^3}\sqrt g  \phi^+\l(\pa_t \hat
P+(\pa_k\al^k+\al^k\pa_k\ln\sqrt g-2\al^k\Up_k)\hat P+i[\H,\hat P]\r)\phi d^3x, \nn\\
&=&\Re\int_{S^3}\sqrt g \phi^+\l(\pa_t \hat P+i[\H,\hat P]\r)\phi
d^3x, \label{4.12}
 \end{eqnarray}
 where any Hermitian operator $\hat P$ means $P=\int_{S^3}\sqrt g \phi^+\hat P\phi
 d^3x$ is real for any $\phi$. (\ref{4.12}) clearly shows the connection $\Up^\mu$ has only geometrical
effect, which cancels the derivatives of $\sqrt g$. Obviously, we
cannot get (\ref{4.12}) from   conventional definition of spinor
connection $(\Ga_\mu,\wt\Ga_\mu)$.

Define   4-dimensional momentum of the spinor by
\begin{eqnarray}
p^\mu=\Re \int_{S^3}\phi^+\hat p^\mu \phi\sqrt{g} d^3x. \label{4.10}
 \end{eqnarray}
For a spinor at energy eigenstate, we have   classical approximation
$p^\mu = m u^\mu$, where $m$ defines the classical mass of the
spinor. Let $\hat P=\hat p_\mu$,  we get classical approximation as
$q^\mu\to v^\mu\dl^3(\vec x-\vec X)$,
\begin{eqnarray}
\frac d {dt} p_\mu &=& \Re\int_{S^3}\l(e(\pa_\mu A_\nu-\pa_\nu
A_\mu) q^\nu +\phi^+ \pa_\mu(\Om_\nu \hat s^\nu)\phi-\phi^+(\pa_\mu
\al^\nu)\hat p_\nu\phi\r)\sqrt g d^3x.
\nn\\
&\to & \l[e(\pa_\mu A_\nu-\pa_\nu A_\mu) u^\nu+ s^a \pa_\mu \om_a
\r] \sqrt{1-\bar g_{kl}v^k v^l} -K_\mu,
\label{4.14}\\
s^a &  = &  \Re\int_{S^3}\phi^+ \hat s^a \phi \sqrt {g}
d^3x/\sqrt{1-\bar g_{kl}v^k v^l}=L^a_{~b} \bar s^b,
\label{spn}\\
K_\mu &  =& \Re\int_{S^3}\phi^+(\pa_\mu \al^\nu)\hat p_\nu\phi\sqrt
g d^3x,\label{4.15}\end{eqnarray} in which $\bar
s^b=\int_{R^3}\phi^+\hat s^b\phi d^3X$ is   proper spin of the
spinor. $\bar s^b$ equals to $\pm \frac 1 2 \hbar$ in one direction
but vanishes in other directions. $L^a_{~b}$ is the local Lorentz
transformation between   local tetrad and the central coordinate
system of the spinor\cite{mass}.

Now we prove the following classical approximation of $K_\mu$,
\begin{eqnarray}
K_\mu &\to & g_{\mu\nu}\Ga^\nu_{\al\be} p^\al u^\be \sqrt{1-\bar
g_{kl}v^k
v^l}- p^\nu \frac {d g_{\mu\nu}}{dt} \label{kmu} \\
 &=& \l(\frac 1 2
m ( \pa_\al g_{\mu\be}+\pa_\be g_{\mu\al}-\pa_\mu g_{\al\be}) u^\al
u^\be  - m u^\nu u^\al\pa_\al g_{\mu\nu}\r) \sqrt{1-\bar g_{kl}v^k
v^l}\nn\\
&=&-\frac 1 2 (\pa_\mu g_{\al\be}) m u^\al u^\be \sqrt{1-\bar
g_{kl}v^k v^l}. \label{kmu1}
\end{eqnarray}
in which we used $\frac d {d\tau}=u^\al\pa_\al$. (\ref{kmu1}) can be
proved by using Theorem 4 in \cite{eng} as follows. In this case we
have $\al^\nu=h^\nu_{~a}\bar \al^a$, where $\bar \al^a$ is matrix in
Minkowshi space-time. By Theorem 4 we have
\begin{eqnarray}
\frac {\pa h^\nu_{~a}}{\pa g_{\al\be}} &=&-\frac 1 4
(h^\al_{~a}g^{\nu\be}+h^\be_{~a} g^{\al\nu})-\frac 1 2
S^{\al\be}_{ab}h^\nu_{~n}\eta^{nb}, \label{dhg}\\
S^{\al\be}_{ab} &=& \frac 1 2
(h^\al_{~a}h^\be_{~b}+h^\be_{~a}h^\al_{~b})\mbox{sgn}(a-b).
\label{sdf}\end{eqnarray} Then we get
\begin{eqnarray}
(\pa_\mu \al^\nu)\hat p_\nu  &=& \pa_\mu g_{\al\be} \frac{\pa
h^\nu_{~a}}{\pa g_{\al\be}} \bar \al^a p_\nu = \pa_\mu g_{\al\be}
\l(-\frac 1 4(\al^\al p^\be +\al^\be p^\al)-\frac 1 2
S^{\al\be}_{ab}h^\nu_{~n}\eta^{nb}\bar \al^a
p_\nu\r)\nn\\
&=& -\frac 1 4 \pa_\mu g_{\al\be} \l((\al^\al p^\be +\al^\be p^\al)+
2 S^{\al\be}_{ab}h^\nu_{~n}\eta^{nb}\bar \al^a p_\nu\r).
\label{sdf0}
\end{eqnarray}
For classical approximation we have
\begin{eqnarray}
\phi^+\bar \al^a\phi \to \bar v^a\dl^3(\vec x -\vec X), \quad
h^\nu_{~n}\eta^{nb} p_\nu\phi \to m\bar u ^b\phi,\quad
S^{\al\be}_{ab}=-S^{\al\be}_{ba}.\label{appr}\end{eqnarray}
Substituting (\ref{sdf0}) and (\ref{appr}) into (\ref{4.15}), we get
the right hand term of (\ref{kmu1}).  The proof is finished.

Substituting (\ref{kmu}) into (\ref{4.14}) and noticing $d\tau =
\sqrt{1-\bar g_{kl}v^k v^l } dt$, we get   Newtonian second law for
the spinor
\begin{eqnarray}
\frac {d } {d\tau}p^\mu+\Ga^\mu_{\al\be} p^\al u^\be=  g^{\al
\mu}\l(e(\pa_\al A_\be-\pa_\be A_\al) u^\be+s^a\pa_\al \om_a\r).
\label{gds1}
\end{eqnarray}
Although we derive (\ref{gds1}) in GCS, it obviously holds in all
coordinate system due to the covariant form. Clearly, the additional
acceleration in (\ref{gds1}) is a little different from (\ref{acc}).
If the spin-gravity coupling potential $s_\mu\Om^\mu$ can be
ignored, (\ref{gds1}) satisfies   `mass shell constraint' $\frac d
{d t}(p^\mu p_\mu) =0$\cite{gu2,mass,masshell}. In this case, the
classical mass of the spinor is a constant and the free spinor moves
along geodesic.

In (\ref{gds1}) we get a spin-gravity coupling potential
\begin{eqnarray}
\Psi\equiv \om_a s^a=\Om_\al s^\al.\label{sgp}
\end{eqnarray}
This potential provides an explanation for the relevance between
magnetic field and rotation of a celestial body. For a static star
without rotation, the magnetic field is also very weak, because in
this case we have $\Om_\mu=0$ and the spins of all particles have
not a dominant direction, and their magnetic fields are canceled
each other. In a rotational star, we have $\Om_\mu\ne 0$, and the
spins are automatically arranged in order to generate macro magnetic
field. This macro magnetic field is in turn enhanced by the orbital
magnetic moment of particles.

For many body problem,   dynamics of the system should be juxtaposed
(\ref{4.3*}) due to the superposition of Lagrangian,
\begin{eqnarray} i(\pa_t+\Up_t)\phi_n=\H_n\phi_n,\qquad \H_n= -\al^k \hat p_k+eA_0+m_n \ga_0-\Om_\mu \hat s^\mu. \label{hamlt}
\end{eqnarray}
The coordinate, speed and momentum of $n$-th spinor are defined
by\cite{gu2,mass},
\begin{eqnarray}
\vec X_n(t)= \int_{S^3} \vec x q^0_n\sqrt{g} d^3x,\quad \vec
v_n=\frac d {dt} \vec X_n, \quad p^\mu_n=\Re \int_{S^3}\phi^+_n\hat
p^\mu \phi_n\sqrt{g} d^3x.\label{dfn}
\end{eqnarray}
The classical approximation condition for point-particle model
reads,
\begin{eqnarray}
q^\mu_n\to u^\mu_n \sqrt{1-\bar g_{kl}v^k_n v^l_n}\dl^3(\vec x-\vec
X_n),\quad u^\mu_n\equiv \frac {d X^\mu_n}{d\tau}=(1, \vec v_n)/
\sqrt{1-\bar g_{kl}v^k_n v^l_n}.
\end{eqnarray}
Repeating the derivation from (\ref{4.14}) to (\ref{appr}), we get
  classical dynamics for each spinor,
\begin{eqnarray}
\frac {d } {d\tau}p_n^\mu+\Ga^\mu_{\al\be} p_n^\al u_n^\be=  g^{\al
\mu}\l(e_n(\pa_\al A_\be-\pa_\be A_\al) u_n^\be+s_n^a\pa_\al
\om_a\r),~(\forall n). \label{gdsn}
\end{eqnarray}

\section{discussion and conclusion}
\setcounter{equation}{0}

To split the spinor connection into $\Up_\mu$ and $\Om_\mu$ not only
makes   calculation simple, but also highlights their different
physical meanings. $\Up_\mu$ corresponds to   geometrical
calculations, but $\Om_\mu$ has complex form and leads to dynamical
effects. $\Om_\mu$ couples with the spin $s^\mu$ of a spinor, which
provides location and navigation functions for a spinor. In this
representation, the connection only depends on metric but is
independent of Dirac or Pauli matrices, and their classical
approximation is parallel to the speed of spinor.

The new vector $\Om_\mu$ provides an explanation for the origin of
magnetic field of celestial body. In weak gravity, the spin-gravity
coupling energy is a higher order infinitesimal, but in a neutral
star, this term may become dominant. In a diagonal metric we have
$\Om_\mu=0$, and a static planet is usually of very weak magnetic
field. In (\ref{gds1}), the gravitomagnetic force is caused by
Christoffel symbols  $\Ga^\mu_{\al\be}$. In harmonic coordinate
system, the main part of gravitomagnetism has a similar structure of
Maxwell equation system which was derived in \cite{gm1,gm2,gm3}. The
gravitomagnetic potential is equal to $\vec A =
(g^{01},g^{02},g^{03})$, and field intensity $\vec B = \nb \times
\vec A$. The gravitomagnetic field only interacts with speed $\vec
v$ of a particle, but is independent of spin $s^\mu$. This feature
is different from electromagnetic field.

By (\ref{ps}) we find the spin is actually a true 4-d vector, which
is different from angular momentum, the latter is an axial vector.
Besides, $\Om_\mu$ is also irrelevant with gravitomagnetic field. So
this study may be helpful to understand the marvelous structure and
wonderful property of a spinor, as well as subtle interaction
between spinor and space-time.

In conventional classical approximation we usually use inadequate
limitations such as  $\hbar\to 0$.  $c\to \infty$. They are
constants act as units of physical variables. We can only make
approximation such as $v\ll c$ or (\ref{4.6*}) if the mean radius of
a spinor is much less than moving scale. Most paradoxes and puzzles
in physics are caused by such ambiguous statements or overlapping
concepts of different logical systems. A detailed discussion for
such problems in Minkowski space-time is given in \cite{mass,sbt}.
One of purposes of this paper is to show the consistence of general
relativity, quantum mechanics and classical mechanics.

It is a good choice to take Pauli or Dirac matrices as tetrad, and
then the expression of equations and meanings of parameters become
simpler and clearer as shown above. In fact, all current fundamental
physical theories can be simply unified in this elegant language as
follows:

{\bf A1.} {\em The space-time is described by
\begin{eqnarray}
d\mathbf{x}=\wt\ga_\mu dx^\mu=\ga_a \dl X^a,
\label{1.1a}\end{eqnarray} in which  $\ga_a$ and $\wt\ga_\mu$
satisfy the $C\ell({1,3})$ Clifford algebra} (\ref{1.2a}).

{\bf A2.} {\em The dynamics for a definite physical system is given
by
\begin{eqnarray}\pa \Psi ={\cal F}(\Psi), \qquad \pa \equiv \wt \ga^\mu \pa_\mu, \label{1.5a}
\end{eqnarray}
in which $\Psi=(\psi_1,\psi_2,\cdots,\psi_n)^T$, and ${\cal
F}(\Psi)$ consists of some tensorial products of $\Psi$, so that the
total equation is covariant. }

\acknowledgments{I is my pleasure to acknowledge Prof. James M.
Nester for his enlightening discussions and encouragement.  He also
corrected an important mistake in the previous version.}


\begin{thebibliography}{99}

\bibitem{sp1} M. Mathisson, Acta Phys. Pol. 6, 163 (1937).
\bibitem{sp2} A. Papapetrou, Proc. R. Soc. London 209, 248 (1951).
\bibitem{sp3} W.G. Dixon, Phil. Trans. R. Soc. London A277, 59 (1974).
\bibitem{FFF}  P.M. Alsing, G.J. Stephenson Jr, and Patrick Kilian, {\em Spin-induced non-geodesic motion, gyroscopic
precession, Wigner rotation and EPR correlations of massive spin-1/2
particles in a gravitational field}, arXiv:0902.1396
\bibitem{spin1} Yuri N. Obukhov, {\em On gravitational interaction of fermions},
Fortsch. Phys. 50 (2002)711-716, arXiv:gr-qc/0112080
\bibitem{spin2} H. Behera, P. C. Naik, {\em Gravitomagnetic Moments and
Dynamics of Dirac's (spin 1/2) fermions in flat space-time
Maxwellian Gravity},  Int. J. Mod. Phys. A19 (2004)4207-4230,
arXiv:gr-qc/0304084.
\bibitem{spin3} I. B. Khriplovich, A. A. Pomeransky, {\em Gravitational Interaction of
Spinning Bodies, Center-of-Mass Coordinate and Radiation of Compact
Binary Systems}, Phys.Lett. A216 (1996)7, arXiv:gr-qc/9602004.
\bibitem{spin4} B. Mashhoon, D. Singh, {\em Dynamics of extended spinning masses in a gravitational
field}, Phys. Rev. D74, 124006 (2006).
\bibitem{spin5} B. Mashhoon, {\em Neutron interferometry in a rotating frame
of reference}, Phys. Rev. Lett. 61, 2639-2642 (1988).
\bibitem{spin6} F. W. Hehl, W. T. Ni, {\em Inertial effects of a Dirac particle},
Phys. Rev. D 42, 2045-2048 (1990).
\bibitem{spin7} B. J. Venema, P. K. Majumder, S. K. Lamoreaux, B. R. Heckel
and E. N. Fortson, {\em Search for a coupling of the Earth's
gravitational field to nuclear spins in atomic mercury}, Phys. Rev.
Lett. 68, 135-138 (1992).
\bibitem{LT1} H. Thirring, {\it On the effect of
rotating distant masses in Einstein's theory of gravitation}, Gen.
Rel. Grav. {\bf 16} (1984) 712-725
\bibitem{LT2} J. Lense and H. Thirring,  {\it On the influence of the proper
rotation of central bodies on the motions of planets and moons
according to Einstein's theory of gravitation}, Gen. Rel. Grav. {\bf
16} (1984) 727-750.
\bibitem{Schiff1} L. I. Schiff, {\it On experimental tests of the general theory of
relativity}, {Am. J. Phys.} {28} (1960) 340-343.
\bibitem{Schiff2} L. I. Schiff, {\it Motion of gyroscope according to Einstein's
theory of gravitation}, {Proc. Nat. Acad. Sci.} {46} (1960) 871-882
\bibitem{Schiff3} L. I. Schiff, {\it Possible new experimental test of general
relativity theory}, {Phys. Rev. Lett.} {4} (1960) 215-217.
\bibitem{EEP9}  K. Nordtvedt, {\em Lunar Laser Ranging - A Comprehensive Probe of
Post-Newtonian Gravity}, arXiv:gr-qc/0301024
\bibitem{gyro2} K. Nordtvedt, {\em Existence of the gravitomagnetic interaction},
Int. J. of Theo. Phys., 27, 1395-1404 (1988).
\bibitem{gyro4} C. M. Will, {\em Theory and Experiment in Gravitational Physics},
Cambridge University Press, New York, (1993)
\bibitem{gm1} B. Mashhoon, {\em Time-Varying Gravitomagnetism}, Class. Quant. Grav. 25:085014(2008), arXiv:0802.1356
\bibitem{gm2} D. Bini, C. Cherubini, C. Chicone, B. Mashhoon, {\em Gravitational induction},
Class. Quant. Grav. 25:225014(2008), arXiv:0803.0390
\bibitem{gm4} B. Mashhoon, {\em Gravitoelectromagnetism: A Brief
Review,  the third chapter of The Measurement of Gravitomagnetism: A
Challenging Enterprise}, edited by L. Iorio (Nova Science, New York,
2007), pp. 29-39, arXiv:gr-qc/0311030
\bibitem{gyro5} T. W. Murphy, Jr,  K. Nordtvedt and S. G. Turyshev, {\em  The Gravitomagnetic Influence
on Gyroscopes and on the Lunar Orbit}, Phys. Rev. Lett. 98:071102
(2007), arXiv:gr-qc/0702028
\bibitem{prec} Rickard Jonsson,  {\em A covariant formalism of spin
precession with respect to a reference congruence}, reference:
Class. Quant. Grav. 23:37-59 (2006), arXiv:0708.2533
\bibitem{1} M. Sachs, {\em General relativity and matter} {\bf (Ch.3)}, D. Reidel, 1982.
\bibitem{2} W. L. Bade, H. Jehle, Rev. Mod. Phys.{\bf  25(3)}, (1953)714
\bibitem{3} P. G. Bergmann, Phys. Rev. {\bf 107(2)}, (1957)624
\bibitem{4} D. R. Brill, J. A. Wheeler, Rev. Mod. Phys. {\bf 29(3)}, (1957)465
\bibitem{5} J. P. Crawford, Adv. Appl. Cliff. Alg. {\bf 2(1)}, (1992)75
\bibitem{eng} Y. Q. Gu, {\em The Vierbein Formalism and Energy-Momentum Tensor of Spinors}, arXiv:gr-qc/0612106
\bibitem{nst} J. M. Nester, Journal of Mathematical Physics 33, 910 (1992).
\bibitem{dmh} A. Dimakis and F. Muller-Hoissen, Class. Quantum Grav. 8, 2093
(1991).
\bibitem{gu2} Y. Q. Gu, {\em New Approach to N-body Relativistic Quantum Mechanics}, Int. J. Mod. Phys. A22:2007-2020(2007),
arXiv:hep-th/0610153
\bibitem{mass} Y. Q. Gu, {\em Local Lorentz Transformation and Mass-Energy Relation of Spinor},  arXiv:hep-th/0701030, accepted by physical
essays
\bibitem{masshell} P. M. Alsing, J. C. Evans and K. K. Nandi, {\em The phase of a quantum mechanical particle in
curved spacetime}, Gen. Rel.  Grav. 33, 1459-1487 (2001); gr-qc/
0010065.
\bibitem{gm3} Y. Q. Gu, {\em Stationary Spiral Structure and Collective Motion of the Stars in a Spiral Galaxy},
arXiv:0805.2828
\bibitem{sbt} Y. Q. Gu, {\em Some Subtle Concepts in Fundamental Physics}, Physics Essays Vol. 30: Pages 356-363(2017), arXiv:0901.0309
\end{thebibliography}
\end{document}